# A COMBINATION OF TEMPORAL SEQUENCE LEARNING AND DATA DESCRIPTION FOR ANOMALY-BASED NIDS

Nguyen Thanh Van[1,2], Tran Ngoc Thinh[1], Le Thanh Sach[1]

[1]Faculty of Computer Science and Engineering. Ho Chi Minh City University of Technology, VNUHCM, VietNam.
[2]Ho Chi Minh City University of Technology and Education, VietNam.

*ABSTRACT*

*Through continuous observation and modelling of normal behavior in networks, Anomaly-based Network Intrusion Detection System (A-NIDS) offers a way to find possible threats via deviation from the normal model. The analysis of network traffic based on time series model has the advantage of exploiting the relationship between packages within network traffic and observing trends of behaviors over a period of time. It will generate new sequences with good features that support anomaly detection in network traffic and provide the ability to detect new attacks. Besides, an anomaly detection technique, which focuses on the normal data and aims to build a description of it, will be an effective technique for anomaly detection in imbalanced data. In this paper, we propose a combination model of Long Short Term Memory (LSTM) architecture for processing time series and a data description Support Vector Data Description (SVDD) for anomaly detection in A-NIDS to obtain the advantages of them. This model helps parameters in LSTM and SVDD are jointly trained with joint optimization method. Our experimental results with KDD99 dataset show that the proposed combined model obtains high performance in intrusion detection, especially DoS and Probe attacks with 98.0% and 99.8%, respectively.*

## 1. INTRODUCTION

The development of computer networks and Internet network has given rise to critical threats such as zero-day vulnerabilities, mobile threats, etc. Despite recent researches about the cyber security have increased significantly, it only mitigates intrusions because of the huge appearance of many new and sophisticated attacks. A-NIDS is very efficient to protect target systems and networks against attacks. The system can find possible threats via deviation from the normal model or classification normal/abnormal; therefore, it has the ability to detect attacks which are new to the system. In anomaly detection, anomalies are very important because they are serious events and maybe attacks which damage computer and network. For example, an unusual traffic pattern in a network could mean that a computer is attacked and data is transmitted to unauthorized destinations. Therefore, the different types of anomaly will have the correlation with the attacks based on the nature of the anomaly and they need to be detected by A-NIDS. Some existing solutions apply classic anomaly detection systems to make decision-based on the traffic features of the present moment. In the network environment, traffic is generated during the data communications over time, so there is a relationship between packets inside the network traffic. Therefore, examining a single network packet based on short-term features will be less effective to detect attacks, especially when attacks spread over many packets such as APTs, DDoS, the time horizon can span from days to minutes and one can also seconds. In this context,

International Journal of Network Security & Its Applications (IJNSA) Vol. 11, No.3, May 2019

using Long Short Term Memory (LSTM) [1] will be believed to have the unique ability to capture long term aspects of a time series. This important point shows that time series can be a good choice for A-NIDS where attacks are launched in the form of the sequence of packets. LSTMs are a neural model for sequential data and they account for long-term dependencies in information sequences. Therefore, it has the ability to learn long-term dependency and context in sequential data, meaning that temporal features in sequences are achieved. With this advantage of LSTM, we propose model uses LSTM for analyzing and learning network data to extract temporal features by exploring the time-dependent structure from the relationship between packets in input sequences.

A challenge in the A-NIDS is the labels for normal network data is usually available while abnormal data is difficult to obtain. To eliminate bias towards majority group, an algorithm-level solution is used to apply one-class learning that focuses on the target group, creating a data description [2]. In data description, normal data is considered to build data description, then it is effectively applied to detect abnormal data or exception points that cannot be matched with this description. The data description (called One-class classification) has many methods such as one-class Support Vector Machine (OC-SVM) [3] and Support Vector Data Description (SVDD) [4]. OC-SVM builds a hyperplane in a feature space that separates the normal data from the origin with maximum margin while SVDD tries to find a sphere with a minimum volume containing all the data objects. One-class classifications have a major limitation that is the lack of ability to handle dynamic systems. This can be addressed by LSTM which converts time series to fixed long vectors before applying OC-SVM or SVDD. In this paper, we propose an approach combining LSTM and One-class classification method SVDD to take advantage and interactive supports from the two strategies for detecting anomalous in network traffic. Particularly, the parameters of the LSTM structure and the SVDD formulation are jointly trained with joint optimization methods. The combination is inspired from the work [5] in 2017 which authors proposed a general combination model and simulated in some datasets such as occupancy, exchange rate, HTTP, and stock price. To the best of our knowledge, the combination model is considered as the first work in the intrusion detection domain.

This paper is organized as follows. Section 2 discusses the background and related works. Section 3 describes our proposed combination model. Section 4 describes our experiments and results. Section 5 concludes.

## 2. BACKGROUND AND RELATED WORKS

### 2.1. TYPE OF NETWORK ANOMALY

In A-NIDS, intrusions are detected based on anomaly detection, therefore, anomaly detection techniques [6] are applicable in the intrusion detection domain. An important aspect of an anomaly detection technique is the nature of the desired anomaly. Anomalies can be classified into the following three categories. The first, if an individual data instance can be considered as anomalous with respect to the rest of the data, then the instance is termed as a point anomaly. Example, local access to get the privilege of sending packets on the network, an attacker who uses the trial and return much time to guess password compared to the normal range of access for that person will be a point anomaly. The second, if a data instance is anomalous in a specific context then it is termed as a contextual anomaly. Suppose an individual usually has daily normal access to network system except in end month day, when it reaches high. A large range of access in a day of the middle month will be considered a contextual anomaly since it does not conform to the normal behaviour of the individual in the context of time. The third, if a collection of related data instances is anomalous with respect to the entire data set, it is termed as a collective anomaly. The individual data instances in a collective anomaly may not be anomalies by themselves, but their occurrence together as a collection is anomalous. In the case of a DoS attack, multiple requests to connect to a web server are a collective anomaly but a single request is normal. Therefore, we can consider DoS attacks as a collective anomaly. Contextual and collective anomalies have been most commonly explored for sequence data. In the computer network domain, network traffic can be defined as large network packet datasets, which are generated during the data communications over time. Therefore, network traffic datasets can be analyzed as a time series. Network anomalies should be considered as contextual and collective anomalies, so the immediate application of countermeasures need to be implemented.

## 2.2. ANOMALY-BASED NIDS TECHNIQUES AND PREVIOUS WORKS

In the past years, several different techniques have been used in anomaly-based NIDS [7] such as statistical-based, knowledge-based, and machine learning-based. Almost all works considered anomaly detection as a classification problem that builds a model of normal network behaviors to detect new patterns that significantly deviate from the model. However, they do not take into account the previous, recent events to learn long-term dependency and context in the network traffic. Thus, challenges need to be scrutinized to improve performance and make suitable solutions with real network data characters. In general terms all of A-NIDS approaches consist of the following basic stages as shown in Figure1 [8]. A-NIDS observes changes in data stream that are collected from network traffic or host activities by building a profile of the system which is being monitored. The profile is generated over a period of time, so network traffic is considered as time series data. Analyzing network traffic based on time series has an advantage of exploiting the relationship between the packets in the network traffic and observing trends of behaving over a period of time. Subsequently, the temporal structures from network traffic are learned to extract features and they are used to build intrusion detection model.

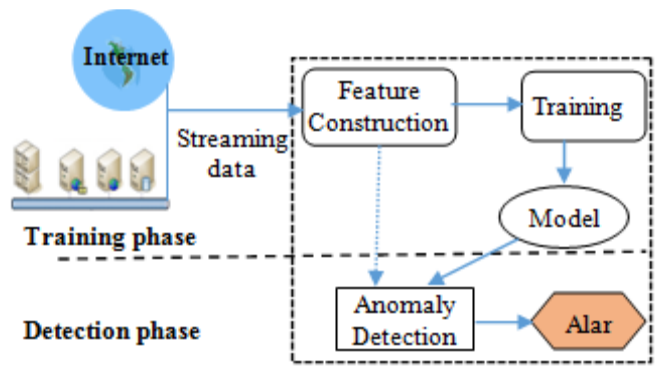

Figure. 1.Anomaly-based NIDS architecture[8]

Recently, the feature-based time series approaches are used to extract the various features from the time series data and combine them to make classifications in time series. However, these approaches are based on handcrafted methods and demand intensive pre-processing work. A few attempts have been made aimed at the application of deep learning approaches for time series processing problems. Deep Learning [9] is used to combine the feature extraction of time series with the non-linear autoregressive model for higher level prediction. The good feature representations are obtained from a large amount of unlabeled data, so the model can be pre-trained in a completely unsupervised fashion. In a previous work [10], we applied Auto Encoder technique to A-NIDS with the goal of learning features automatically and our experimental results show that it is an effective technique for exact intrusion detection with low error rate. As a deep neural network, LSTM is widely used for processing time series data [1], which is an improved model based on Recurrent Neural Networks (RNNs). A more advanced RNN architecture with several control structures, LSTM uses well-designed "gate" structures to decrease the vanishing gradient problem during errors back propagation. Subsequently, the loss can go back-wards through longer time step, which enables LSTM to learn long-term dependency and context. With that advantage, LSTM is used to learn temporal features in many image applications, for instance, Y.Feng [11] proposed a new feature learning method for gait recognition for to preserve temporal information in a gait sequence. Another work [12] utilized the LSTM units' ability to findlong temporal relation from its input sequences as well as extracting local and dense features through convolution operations. However, using LSTM to learn temporal features in network security is rare. LSTM is also applied in classification and anomaly detection. Some researchers have applied LSTM for IDS such as Kim. J., [13] has applied the LSTM along with Gradient Descent Optimization for an effective intrusion detection classifier with an accuracy of 97.54% and recall of 98.95%. Staudemeyer, R. C., [14] evaluated the performance of LSTM networks on the KDD99 IDS data set with satisfactory results. Then they improved results in which the training accuracy 93.82%. These classifications using LSTM need both normal and abnormal data for training, therefore it is inconvenient in many imbalance data applications. In using LSTM for detecting anomalous in computer network, Loıc Bontemps et al. [15] trained LSTM RNN with normal time series data before performing a live prediction for each time step. The model is built on a time series version of the KDD99 dataset. Their experiments demonstrate that it is possible to offer reliability and efficiency for collective anomaly detection. Min Ch. in [16] used a multi-scale LSTM model to detect anomalous Border Gateway Protocol (BGP) traffic by considering the Internet flow as a multi-dimensional time sequence and learn the traffic pattern from historical features in a sliding time window. Their obtained highest result is 99.5% with window size 30 when detecting Slammer anomaly. LSTM is used to anomalous detect in time series for many domains [17].

Similar to most neural networks, in LSTM, there is the employment of the Softmax function as its final output layer for its prediction, and the cross-entropy function for computing its loss. With that way, they do not directly optimize an objective criterion for anomaly detection, which results in challenges to optimization problems. To solve this problem, a method is introduced using a combination of LSTM and another model that is specific for anomaly detection. Some works present an amendment to this norm by introducing linear support vector machine (SVM) as the replacement for Softmax in a GRU model such as [18] and they reached a training accuracy of 81.54% and a testing accuracy of 84.15%. This work and most of the current researches on anomaly detection are based on the learning of normally and anomaly behaviors (binary classification). However, in many real cases in anomaly detection related applications, normal examples are available, while the abnormal data are rare or difficult to obtain. Therefore, anomaly detection techniques cannot learn both normally and anomaly behaviors and they only focus on the normal data and aim to build a description of them. This description is then applied to detect abnormal data that cannot fit this description very well [19]. Learning from imbalanced data have been mainly driven by many real life applications. In those applications, we can see the minority class is usually the more important one and hence many methods are required to improve its recognition rates. This is closely related to our problem in intrusion detection where normal data is often available while intrusion is rare. In network security, some works used One-class SVM to detect anomaly in network traffic [20] and their result archived 71% accuracy; Wireless Sensor Networks [21] to achieve high detection accuracy and low false alarm rate. In order to consider these problems, we combine LSTM and SVDD to take advantage of two strategies to detect abnormalies in network traffic.

## 3. PROPOSED COMBINING LSTM AND SVDD MODEL IN A-NIDS

### 3.1. PROPOSED COMBINATION ARCHITECTURE

The proposed architecture in Figure 2, LSTM is used to learn the temporal structure from network traffic data and SVDD to describe normal examples. Many authors research combining time series model with another classification algorithm to improve performance for time series classification tasks such as in [5] [18] [22]. These works combine time series model with SVM that is supervised learning model with associated learning algorithms that analyses data used for classification. This classification marks each example as belonging to one or the other of two classes. They just distinguish between two (or more) classes and cannot detect outliers which do not belong to any of the classes. Therefore, these methods can potentially fail in many real cases when the abnormal data will go to null with the increasing diversity of the normal data. To deal with this challenge, we use SVDD as data description because it can naturally detect anomaly among the normal data which are closed by a boundary. SVDD [4] is motivated by the Support Vector Classifier [23]. It obtains a spherically shaped boundary around a dataset and similar to the Support Vector Classifier it can be made flexible by using other kernel functions. We find a hyper sphere enclosing the normal data while leaving the anomalies outside. After that, we find a decision function to determine whether a sequence of packets is anomalous or not based on the described data.

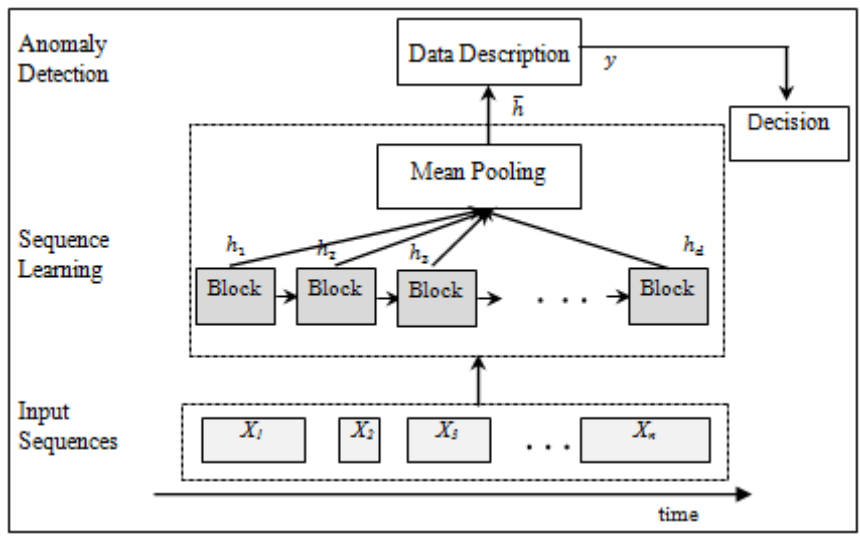

Figure. 2.The proposed combination architecture in A-NIDS.

Every sequence of packets from network traffic data X are fed to LSTM model, each sequence $X_i$ includes several packets and may have different number of packet, they are illustrated like this

$$X = \{X_1, X_2, X_3, X_i \ldots, X_n\}, n: \text{number of sequences.}$$
$$X_i = [x_{i,1} x_{i,2} x_{i,j} \ldots x_{i,d_i}], \quad d_i: \text{number of packets}$$
$$x_{i,j} \in R^p, \forall j \in \{1,2,3 \ldots, d_i\}, \quad p: \text{number of features in a packet}$$

LSTM outputs for each packet sequence are averaged by using mean pooling method. By this way, we can get a new sequence with many new temporal features. After that, new sequences will be input to Data description unit to find a decision function y to determine whether a sequence of packets is anomalous or not based on observed data. The function that takes the value +1 marking a normal sequence and -1 otherwise:

$$y(X_i) = \begin{cases} -1 \text{ if } X_i \text{ is anomaly sequence} \\ +1 \text{ if } X_i \text{ is normal sequence} \end{cases} \quad (1)$$

To find this decision function, we use SVDD to find a small hypersphere enclosing the normal data while leaving the anomalies outside

### 3.2. LSTM Model

This architecture can be seen as a deep architecture through time steps. We assume sequence $i^{th}$ ($X_i$) of packet is fed to LSTM model through time with input vector $x_{i,j} \in R^p$ (the $j^{th}$ LSTM block, $p$ features). Then, we initialize the learning parameters weights and biases with arbitrary values that will be adjusted through training. The cell states $c_{i,j}$ of LSTM are computed based on the input vector $x_{i,j}$ and its learning parameters values. With $x_{i,j}$ is input vector of the $i^{th}$ sequence at the time the $j^{th}$, equations in the $j^{th}$ internal LSTM block are operated as follows

Block input: $\quad z_{i,j} = tanh(W^z x_{i,j} + U^z h_{i,j-1} + b^z) \quad (2)$

Input gate: $\quad i_{i,j} = sigmoid(W^i x_{i,j} + U^i h_{i,j-1} + b^i) \quad (3)$

Forget gate: $f_{i,j} = sigmoid(W^f x_{i,j} + U^f h_{i,j-1} + b^f) \quad (4)$

Cell gate: $c_{i,j} = i_{i,j} \odot z_{i,j} + f_{i,j} \odot x_{i,j-1}) \quad (5)$

Output gate: $\quad o_{i,j} = sigmoid(W^o x_{i,j} + U^o h_{i,j-1} + b^o) \quad (6)$

Block output: $\quad h_{i,j} = o_{i,j} \odot tanh(c_{i,j}) \quad (7)$

Here $W(.)$ are rectangular input weight matrices for input ($z$), input gate ($i$), forget gate ($f$) and output gate ($o$); $U(.)$ are square recurrent weight matrices for input ($z$), input gate ($i$), forget gate ($f$) and output gate ($o$). Two point-wise non-linear activation functions: logistic sigmoid for the gates and hyperbolic tangent for the block input and output. Point-wise multiplication of two vectors is denoted by $\odot$. Through LSTM blocks, we compute the average $\bar{h}_i$ of LSTM outputs for a packet sequence $i^{th}$ or other pooling methods, such as the last in [18] [22] and max.

$$\bar{h}_i = \frac{1}{d_i} \sum_{j=1}^{d_i} h_{i,j} \quad (8)$$

Final in sequence learning period is all output vectors $\{\bar{h}_i\}_{i=1}^n$ of sequences are input to the SVDD model. The combination model will use the loss function of SVDD with the introduction of SVDD as its final layer. Therefore, the parameters of the model are also learned by joint optimizing the objective function of SVDD.

### 3.3. SVDD Model

To detect anomalous in every sequence, we use support vector data description (SVDD) that is rewritten in a form comparable to the support vector classifier (SVC). It poses the ability to map the data to a new, high dimensional feature space without any extra computational costs. Therefore, we can obtain more flexible descriptions and it will be presented how the outlier sensitivity can be controlled in a flexible way. To describe normal data using SVDD, we need to find a small hyper sphere with radius R and center c to separate the anomalies from the normal data. Objects on the boundary are the support vectors with $\xi_i = 0$ while outside objects have $\xi_i > 0$.

To minimize a hyper sphere enclosing normal packet sequences, we need to solve an optimization problem that is formulated as follows

$$\min_{\theta,R,c,\xi} R^2 + \frac{1}{n\nu}\sum_i \xi_i \quad (9)$$

$$\text{s.t:} \|\bar{h}_i - c\|^2 \leq R^2 + \xi_i, \xi_i \geq 0, \forall i \quad (10)$$

$$W^{(.)T}W^{(.)} = I,\ U^{(.)T}U^{(.)} = I, b^{(.)T}b^{(.)} = 1. \quad (11)$$

Here hyper sphere with center $c$ and radius $R$; input data $\bar{h}_i$ is the output vector from LSTM model that is calculated in equation (8); $n$: normal examples in training set; $\xi_i$: slack variables to penalize misclassified - accepting some normal data that located in an unsafe area: outside the hypersphere; $\nu$: regularization parameter: the coefficient of controlling trade-off between the size of the hypersphere and the total error $\xi_i$; $\theta$ represents all LSTM parameters. The free parameters, $\theta, R, c, \xi$ have to be optimized, taking the constraints (10-11) into account. The first constraint that almost all objects are within the sphere, it also accepts some normal data located outside of the hyper sphere. The second orthogonality constraints are used to input LSTM's output vectors. They guarantee that input vectors are orthogonal matrices that can effectively prevent the gradient vanishing/explosion problem in conventional LSTM. An optimization algorithm is used for loss minimization and it adjusts the weights and biases based on the computed loss. The trained model can be used for anomaly detection on a given data based on decision function as follows

$$y(X_i) = \text{sgn}\left(R^2 - \|\bar{h}_i - c\|^2\right). \quad (12)$$

### 3.4. Training Method

A discussion in [5] showed that the gradient descent based training method provides higher performance due to its learning capabilities. In our problem, we use a training approach based on only the first order gradients, which updates the parameters at the same time. However, we need to require an approximation to the original SVDD formulation to apply this method. We also ensure the convergence of the approximated formulation to the original SVDD formulation. We study slack variable in constraints (10) that can be incorporated into formula (9). Therefore, we need rewrite the first constrained optimization as a part of objective function. Consequently, the first constraint (10) can be equivalent to

$$\xi_i = \max\{0, \psi_{R,c}(\bar{h}_i)\},$$
$$\text{with } \psi_{R,c}(\bar{h}_i) = \|\bar{h}_i - c\|^2 - R^2. \quad (13)$$

We can write slack variable $\xi_i$ as a function $P(.)$ as follows

$$P\left(\psi_{R,c}(\bar{h}_i)\right) = \max\{0, \psi_{R,c}(\bar{h}_i)\}. \quad (14)$$

By this way, we can eliminate the first constraint (10). The learning problem is equivalent to the constrained optimization problem over R, $c$ and $\theta$ as follows

$$\min_{\theta,R,c} R^2 + \frac{1}{n\nu}\sum_i P\left(\psi_{R,c}(\bar{h}_i)\right) \quad (15)$$

$$W^{(.)T}W^{(.)} = I,\ U^{(.)T}U^{(.)} = I, b^{(.)T}b^{(.)} = 1 \quad (16)$$

Because $P(.)$ is non-differential function, we cannot optimize (15) by using gradient descent algorithm. To deal with this obstacle, we can consider objective function as Hinge loss function and solving it. Another way, we can approximate function $P(.)$ to a function that is differentiable function with smooth parameter $\chi$ [31] in (17)

$$Q_\chi\left(\psi_{R,c}(\bar{h}_i)\right) = \frac{1}{\chi} \ln\left(1 + e^{\chi \psi_{R,c}(\bar{h}_i)}\right) \quad (17)$$

Where $\chi$ is smooth parameter – it is handled to expect that $Q_\chi(.)$ converges to $P(.)$. According (17), $Q_\chi(.)$ converges to $P(.)$ when $\chi$ increases. Thus, in our experimentation, we need to choose a large value for $\chi$. When $Q_\chi\left(\psi_{R,c}(\bar{h}_i)\right)$ converges to $P\left(\psi_{R,c}(\bar{h}_i)\right)$, as a consequence, an approximation $F_\chi(c, R, \theta)$ of SVDD objective function $F(c, R, \theta)$ converges to $F(c, R, \theta)$.
Now, the optimization problem is defined as

$$\min_{\theta, R, c} F_\chi(c, R, \theta) = R^2 + \frac{1}{nv} \sum_i Q_\chi\left(\psi_{R,c}(\bar{h}_i)\right) \quad (18)$$

$$W^{(.)T}W^{(.)} = I, \; U^{(.)T}U^{(.)} = I, b^{(.)T}b^{(.)} = 1 \quad (19)$$

Parameters $c, R, \theta$ are updated till obtaining optimization values for (18) and (19). $F(.)$ should be minimized with respect to $c, R, \theta$. After that, we use gradient descent algorithm to train our combination model makes the parameters in both models LSTM and SVDD ($c, R, \theta$) are jointly optimized. Each iteration $k$ involves cycling through the training data with the updates.

## 4. EXPERIMENTS AND RESULTS

### 4.1. DATASET

KDD99 dataset[24] is widely used as one of the few publicly available datasets for IDS problems. It is believed to apply as an effective benchmark data set to help researchers compare different intrusion detection methods. Although it is still not an accurate real-world dataset, there are many papers that describe their implementations on this dataset specifically. This paper will use the KDD99 dataset, however, our work would work with any dataset that conforms to these rules. There are many recent datasets containing more modern attacks, such as the UNSW-NB15 dataset generated for the Australian Centre for Cyber Security [25], Intrusion Detection Evaluation Dataset (CICIDS2017) contains benign and the most up-to-date common attacks [26], Unified Host and Network Dataset[27] collected from the Los Alamos National Laboratory enterprise network over the course of approximately 90 days. Most of these datasets may be more applicable and newer for recent use cases, however, they are not used as publicly as KDD99.

The KDD99 dataset consists of approximately 4,900,000 single connection records. The distribution of data samples is presented in Table 1. This dataset must deal with the imbalanced data problem. All data is pre-processed to match to the detection methods proposed.

Table 1. KDD99 dataset

| Dataset | Normal | Abnormal | | | | Total |
|---|---|---|---|---|---|---|
| | | *DoS* | *Probing* | *R2L* | *U2R* | |
| All KDD99 | 972,780 | 3,883,370 | 41,102 | 1,126 | 52 | 4,898,430 |
| | 19.859% | 79.278% | 0.839% | 0.023% | 0.001% | 100% |
| Test KDD99 | 60,593 | 229,853 | 4,166 | 16,189 | 228 | 311,029 |
| | 19.481% | 73.901% | 1.339% | 5.205% | 0.073% | 100% |

## 4.2. EXPERIMENTS USING COMBINATION MODELS

We experiment different combined models, ANN+SVDD and LSTM+SVDD with some configurations such as 256/1024 nodes and 5/10 time steps look-back on the same dataset: ANN_256+SVDD, ANN_1024+SVDD, LSTM_5+SVDD and LSTM_10+SVDD. We also perform single models: LSTM and SVDD to compare them with combined models. To evaluate single SVDD model, we take every packet as a sequence of input to SVDD. Then, we input normal data to train with a gradient descent algorithm to optimize the parameters. For single LSTM model, we employ the conventional Softmax function as the final output layer for the prediction of LSTM structure, and the cross-entropy function for computing its loss. In ANN+SVDD model, we combine a neural network with a data description model. The model is trained using gradient descent algorithm to optimize the parameters. To evaluate the performance of the combination model, LSTM+SVDD, we use an LSTM cell with many blocks as a hidden layer in regarding learn time-based features, then the loss function of SVDD with the attendance of SVDD as its final layer. Therefore, the parameters of the model are also learned by joint optimizing the objective function of SVDD. Every sequence that has different number packet based on choosing different look-back parameters is input to LSTM structure. When sequences are input to the model, thanks to LSTM structure that model is able to process such sequences to get fixed length vector and support for SVDD that cannot directly process these sequences. It also automatically learns the time-dependent features, therefore, it helps network traffic data is more completely described. Attack types accuracy are executed by matching actual and predicted attacks are the correct predictions in all experiments.

## 4.3. RESULTS AND DISCUSSION

Table 2. Results of the experiment models

| Methods | Normal | Abnormal | | | | w-sum |
|---|---|---|---|---|---|---|
| | | DoS | Probing | R2L | U2R | |
| SVDD | 83.00 | 97.00 | 95.60 | 8.10 | 81.00 | 94.02 |
| LSTM | 98.99 | 96.55 | 56.14 | 0 | 0.06 | 60.14 |
| ANN_256+SVDD | 88.50 | 97.00 | 99.78 | 29.60 | 92.00 | 97.99 |
| ANN_1024+SVDD | 88.85 | 97.81 | 99.73 | 11.15 | 92.86 | 97.76 |
| LSTM_5+SVDD | 92.90 | 97.00 | 97.50 | 80.00 | 11.00 | 95.91 |
| LSTM_10+SVDD | 96.00 | 98.00 | 99.80 | 86.00 | 52.00 | 98.59 |

The results in Table 2 show that the combined models outperformed with higher overall accuracy than our single models. The proposed combination models have high accuracy detection trend in Normal, Dos, Probe whereas both U2L and R2L have low figures. Especially, the LSTM single model has the lowest in detecting both U2L and R2L attacks. This is because less percentage of class occurrences are available for these attacks for training. However, combining SVDD to LSTM significant improves the accuracy detection percentage of R2L and U2L attacks. To compare our models, the overall performances are weighted computed as a w-sum measurement (based on different distribution of data samples). In all experiments, LSTM_10 +SVDD model is the best with the overall detection accuracy in Normal, Dos and Probe. We also measure the performance of the best combination model with some recent works in detecting various attacks as in Table 3.

The Table 3 shows that detection accuracy of the proposed method in Probe attacks is the highest in all methods. The proposed method is better than almost all methods in detecting DoS, Probe

International Journal of Network Security & Its Applications (IJNSA) Vol. 11, No.3, May 2019attacks. The DoS attack characteristics match with the collective anomaly. Probe attacks are based on a specific purpose to get information and reconnaissance, so they can be matched to contextual anomalies. It is seen that the proposed model has highly effective in detecting attacks that are spread over packets in a period of time by investigating a sequence of packets. The only drawback of the proposed method is the performance for U2R and R2L classes where their values are somewhat on the lower side as compared to other methods in [28] [29] [30].

Table 3. Accuracy comparison with some recent works

| Methods | Year | Normal | Abnormal | | | |
|---|---|---|---|---|---|---|
| | | | DoS | Probing | R2L | U2R |
| LSTM [13] | 2016 | 95.53 | 97.87 | 54.71 | 0.00 | 57.83 |
| LSTM [14] | 2015 | 99.50 | 99.30 | 75.80 | 17.10 | 0.10 |
| Genetic [28] | 2012 | 99.50 | 97.00 | 78.00 | 11.40 | 5.60 |
| SVM [29] | 2015 | 76.18 | 93.98 | 96.32 | 98.58 | 84.55 |
| KNN [29] | 2015 | 81.17 | 97.63 | 96.27 | 99.67 | 83.24 |
| SVM [30] | 2012 | 99.50 | 97.67 | 91.45 | 53.84 | 90.34 |
| LS SVM (all features) [31] | 2011 | 99.00 | 84.30 | 86.15 | 99.46 | 98.82 |
| Clustering &SVM [32] | 2010 | 99.30 | 99.50 | 97.50 | 19.70 | 28.80 |
| ANN (all features) [33] | 2017 | 88.90 | 99.90 | 98.40 | 42.90 | 87.50 |
| Proposed | 2019 | 96.00 | 98.00 | 99.80 | 86.00 | 52.00 |

We also compare our works to benchmarks of Nicholas J. Miller [34]. Authors used NSL-KDD dataset - is a subset of the KDD99 dataset, so they have the same data characteristic (attack types). Through Table 4, it is seen that our work is better other work in the benchmarks.

Table 4. Accuracy comparison with the benchmarks in [34]

| Algorithm | Accuracy (Total) | Probe | DoS | U2R | R2L | Normal |
|---|---|---|---|---|---|---|
| Naive Bayes | 75.36 | 82.78 | 77.53 | 64.50 | 2.87 | 92.64 |
| Neural Network | 77.80 | 86.10 | 77.45 | 53.02 | 13.21 | 94.84 |
| SVM | 76.91 | 93.89 | 76.08 | 39.00 | 10.78 | 92.84 |
| K-means | 74.04 | 74.19 | 68.73 | 56.01 | 1.23 | 99.09 |
| Our work | 86.36 | 99.80 | 98.00 | 52.00 | 86.00 | 96.00 |

We consider the low performance of U2R and R2L attack types. Both attacks are rare in the dataset and an individual data instance can be considered as anomalous with respect to the rest of the data which are the normal accesses. User to Root (U2R) attacks is illegal to access to the administrator account, exploiting one or several vulnerabilities. Remote to local (R2L) attacks are local access to get the privilege to send packets on the network, the attacker uses trial and error to guess the password. Both U2R and R2L attacks are condition specific and sophisticated. Initiation of these attacks is not similar as compared to others, therefore, these attacks are considered as point anomaly. Consequently, the root problem is coming from LSTM model itself which is highly effective in detecting attacks that are spread over packets in a period while attacks considered as point anomaly are not good. To deal with this challenge, in the near work we will explore the SVDD model. The performance of SVDD can be improved when a few attacks are available. When examples (objects which should be rejected) are available, they can be incorporated in the training to improve the description. In contrast with the training normal examples which should be within the sphere, the attack examples should be outside it. Overall, it can be concluded that the proposed method is quite good considering its performances across both attack and Normal classes, especially attacks have a high frequency such as DoS and Probe. A convenient in our combination models is that we only train normal data to obtain a flexible data description model and it will be used to know how the outlier sensitivity can be controlled in a flexible way. This makes sense in a real dataset where normal examples are available, while the abnormal data are rare or difficult to obtain.

## 5. CONCLUSIONS

In this paper, we propose an approach combining LSTM and a data description model to take advantages and interactive supports from the two strategies to detect anomalous in network traffic. LSTM is candidate of processing time series and supports for obtaining good features from the relationship between packets in a sequence. Using an unsupervised method as SVDD model deal to the high cost of obtaining accurate labels in almost all real application and it is also a good solution for anomaly detection in imbalance data. We apply the gradient-based training method with adjustment the original objective criteria of the combination model to its approximation. The combination gives a high overall performance for A-NIDS and convenient for processing several real datasets. In the future, we will improve processing data, explore more SVDD in order to increase the proportion of accuracy detection both U2L and R2L attacks.

## REFERENCES

[1] Klaus G, Rupesh K. S., Jan K. et al., "LSTM - A Search Space Odyssey", Transactions on neural networks and learning systems, 2017.

[2] Krawczyk and Bartosz, "Learning from imbalanced data: open challenges and future directions," Prog Artif Intell5:221–232, Springerlink.com, 2016.

[3] B. Scholkopf, J. C. Platt, J. Shawe-Taylor et al., "Estimating the support of a high-dimensional distribution," 2001.

[4] D. M. Tax and R. P. Duin, "Support vector data description," in Machine Learning, 2004.

[5] Tolga Ergen, et al."Unsupervised and Semi-supervised Anomaly Detection with LSTM Neural Networks", arXiv:1710.09207 [eess.SP], 2017.

[6] Chandola V., Banerjee A. and Kumar V., "Anomaly detection: A survey," Technical report, USA, 2009.

[7] M. Ahmed, A. Naser Mahmood and J. Hu, "A survey of network anomaly detection techniques," Journal of Network and Computer Applications, p. 13, 2015.

[8] Nguyen Thanh Van and Tran Ngoc Thinh, "Accelerating anomaly-based IDS using neural network on GPU," in IEEE International Conference on Advanced Computing and Applications, 2015.

[9] L. Arnold, S. Rebecchi, S. Chevallier et al., "An Introduction to Deep Learning," in European Symposium on Artificial Neural Networks, Bruges (Belgium), 2011.

[10] Nguyen Thanh Van, Le Thanh Sach and Tran Ngoc Thinh, "An anomaly-based Network Intrusion Detection System using Deep learning," in IEEE International Conference on System Science and Engineering, 2017.

[11] Y. Feng, Y. Li and J. Luo, "Learning Effective Gait Features Using LSTM," in 23rd International Conference on Pattern Recognition (ICPR), México, 2016.

[12] Z. Xu, S. Li and W. Deng, "Learning Temporal Features Using LSTM-CNN Architecture for Face Anti-spoofing," in 3rd IAPR Asian Conference on Pattern Recognition, 2015

[13] Ji K., Jae K., Huong LTT et al., "LSTM - RNN Classifier for Intrusion Detection," in International Conference Platform Technology and Service (PlatCon), South Korea, 2016.

[14] Ralf C. Staudemeyer, "Applying LSTM RNN to intrusion detection," South African Computer Journal, p. 6, 2015.

[15] Lo¨ıc B., Van Cao, James M. et al., "Collective Anomaly Detection based on LSTM RNN," in International Conference on Future Data and Security Engineering, 2016.

[16] Min Ch., Qi.X., J.L. et al., "MS-LSTM: a Multi-Scale LSTM Model for BGP anomaly detection," in 24th International Conference on Network Protocols (ICNP), 2016.

[17] P. Malhotra, L. Vig, G. Shroff et al., "Long Short Term Memory Networks for Anomaly Detection in Time Series," in Presses universitaires de Louvain, 2015.

[18] Agarap and A Fred, "A NN Architecture Combining GRU and SVM for Intrusion Detection in network traffic data," in Machine Learning and Computing (ICMLC), 2018.

[19] Mary H. and Yashwant P. S., "One-class SVM approach to anomaly detection," Taylor & Francis Group, LLC, 2013.

[20] QA Tran, H. Duan and X. Li, "One-class SVM for Anomaly Network Traffic Detection," Researchgate, 2004.

[21] Yang Z., Nirvana M. and Paul H., "Adaptive and Online One-Class SVM-based Outlier Detection Techniques for Wireless Sensor Networks," in Advanced Information Networking and Applications Workshops., 2009.

[22] Abdulrahman A. and Leslie S. S., "A Novel Approach Combining RNN and SVM for time series," in 9th Innovations in Information Technology (IIT), UK, 2013.

[23] Vapnik, "Statistical Learning Theory.," Wiley, 1995.

[24] M. Lincoln, http://kdd.ics.uci.edu/databases/kddcup99.

[25] N. Moustafa and J. Slay, "The evaluation of Network Anomaly Detection Systems: Statistical analysis of the UNSW-NB15 dataset and the comparison with the KDD99 dataset," Information Security: A Global Perspective, pp. 1-14, 2016.

[26] I. Sharafaldin, A. Habibi Lashkari, and Ali A. Ghorbani, "Toward Generating a New Intrusion Detection Dataset and Intrusion Traffic Characterization," in 4th International Conference on Information Systems Security and Privacy (ICISSP), Portugal, Jan, 2018.

[27] M. Turcotte, A. Kent and C. Hash, "Unified Host and Network Data Set," Data Science for Cyber-Security, pp. 1-22, Nov, 2018.

[28] Badran, Khaled, Rockett et al., "Multi-class pattern classification using single, multi-dimensional feature-space feature extraction evolved by multi-objective genetic programming and its application to NID," Genet Program Evolvable, p. 31, 2012.

[29] Abdulla Amin A. and Mamun Bin I. R., "A novel SVM-kNN-PSO ensemble method for intrusion detection system," Applied Soft Computing. © 2015 Published by Elsevier B.V, p. 13, 2015.

[30] Li Y., Xia J., Zhang S. et al., "An efficient intrusion detection system based on support vector machine and gradually features removal method," Expert System with Applications, vol. 39, no. 424–430, p. 7, 2012.

[31] Amiri F., Yousefi M. M. R., Lucas C. et al, "Mutual information based feature selection for intrusion detection. Network and Computer Application,," Network and Computer Applications, vol. 34, no. 1184–1199, p. 16, 2011.

[32] Horng S. J., Su M.-Y., Chen Y. H. et al., "A novel intrusion detection system based on hierarchical clustering and support vector machines.," Expert Systems with Applications, vol. 306–313, p. 8, 2010.

[33] Akashdeep, I. Manzoor and N. Kumar, "A feature reduced intrusion detection system using ANN classifier," Expert Systems With Applications, vol. 88, no. 249–257, p. 9, 2017.

[34] Nicholas J. Miller and Mehrdad Aliasgari, "Benchmarks for evaluating anomaly-based intrusion detection solutions," International Journal of Network Security & Its Applications (IJNSA), vol. 10, no. 5, p. 12, September 2018.